# SCATTERING MATRIX SIMULATIONS OF FIELDS AND DISPERSION RELATIONS IN SUPERCONDUCTING CAVITIES FOR XFEL AND ILC

I. Shinton and R.M. Jones; Cockcroft Institute, Daresbury, WA4 4AD, UK;
University of Manchester, Manchester, M13 9PL, UK.

*Abstract*

The globalised cascaded scattering matrix technique is a well proven, practical method that can be used to simulate large accelerating RF structures in which realistic fabrication errors to be incorporated in an efficient manner without the necessity to re-mesh the entire geometry. The globalised scattering matrix (GMS) technique allows one to obtain the scattering matrix for a structure. The method allows rapid e.m. field calculations to be obtained. Results are presented on monopole mode fields and dispersion relations calculated from direct and analytical methods. Analytical approximate results are also presented for the equivalent shunt susceptance of an iris loaded structure.

## INTRODUCTION

In large accelerating structures such as the ILC beam break up and emmitance dilution are major design concerns; hence the need to be able to accurately model large fractions of these structures in which effects such as wakefields, trapped modes, coupler kicks have been taken into consideration. GSM has been shown in previous works [1-3] to be capable of being employed to accurately model such large scale structures.

## MODE MATCHING USING GSM

Mode matching is an established technique [4-5] which has been applied to various accelerator problems for sharp transitions consisting of adjoining wide narrow (WN) sections [6-10]. The mode matching technique relies upon splitting the structure into a series of sub-regions (WN or NW regions) in which an analytical solution of Maxwell's equations is given in terms of a series expansion over a set of orthogonal modes. The field solutions are then obtained by matching the field at the interfaces. In principle there are an infinite number of modes excited at a junction. In a practical application of the technique we truncate the series with an appropriate ratio of the modes in the narrow to wide section. Indeed, numerous studies have been preformed on the relative convergence phenomena to determine the optimal ratio of modes [11].

## A MODE MATCHING GSM TECHNIQUE FOR FIELD DETERMINATION

The mode matching procedure presented here differs from previous studies. We will consider the case for a propagating monopole mode launched form one port. We derive an analytical relation for the S matrix for a WN or a NW junction. Let us consider the circular WN junction with n modes in region I (wide) and m modes in region II (narrow) where m<=n. We shall denote" b" as the radius of the wide region and "a" as the radius of the narrow region. Analytical electromagnetic fields $\bar{e}$ in a circular beam waveguide are provided in [12]. The characteristic scalar product is given by $\bar{a}_{nm} = \int_S e_n \cdot \hat{e}_m \partial\Gamma$ (over the aperture S) and this evaluates to:

$$\bar{a}_{nm} = \frac{2a^2\chi_n J_0(\chi_{0n}a/b)}{b^2 J_1(\chi_{0n})\left((\chi_{0n}a/b)^2 - \chi_m{}^2\right)} \quad (1)$$

where $\chi_{0n}$ is the nth root of $J_0(\chi_{0n})$. For a = b the normalisation in [12] corresponds to $\bar{a}_{nm} = \delta_{nm}$ (where δ is the Kronecker delta function). The S matrix at the transition is obtained by mode-matching the fields in terms of the modal amplitudes at each transition:

$$S_{21} = 2\left(U + \hat{Z}\bar{a}^t Y \bar{a}\right)^{-1} \hat{Z}\bar{a}^t Y \quad (2)$$

$$S_{11} = \bar{a}S_{21} - U \quad (3)$$

$$S_{22} = 2\left(U + \hat{Z}\bar{a}^t Y \bar{a}\right)^{-1} - U \quad (4)$$

$$S_{12} = \bar{a}\left(U + S_{22}\right) \quad (5)$$

In the above equations U, $Z = Y^{-1}$ is the identity and impedance matrix, respectively. Using Eqs 1-5 the S matrix of a structure decomposed into WN and NW transitions can be obtained. The e.m field in any structure can be subdivided into one of three different regions, as depicted in Fig. 1. Region 1 lies between z=0 and z=1, here the S matrix consists of the GSM formed z=z1 and z=z3.

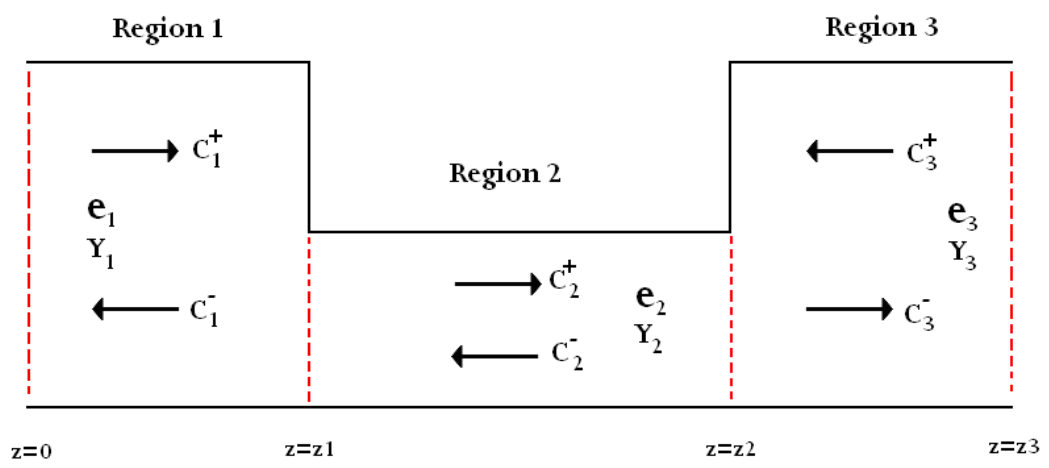

Figure 1: Sketch of WNW transition illustrating incident and reflected modes.

Any subsection of a structure will be classified as a region 2 section, only the first and last sections are classified as region 1 and 3 respectively. The S matrix of any region 2 subsection consists of a propagating matrix representing the infinite propagating wave series $S_{21}^0$ and a reflected matrix $S_{11}^0$ representing the infinite reflected wave series both which are decaying within the subsection [3]. Here the superscripts I and II represent the GSM's left and right of the subsection respectively, $T_z = \delta_{nm} e^{-ik_n z}$ and g is the length of the subsection. Region 3 lies between z=z2 and z=z3, here the S matrix consists of the GSM between z=0 and z=z2. Since the S matrices are directly proportional to the modal amplitudes then after applying the mode matching procedure we obtain the following formulae for the three regions:

<u>Region 1 electric field</u>

$$E_{\parallel} = e_{n1} e^{ik_{n1}(z-z1)} - \sum_{n1=1}^{N} S_{11}(n1,1)\, \bar{e}_{n1} e^{ik_{n1}(z-z1)} Y_{n1} \quad (6)$$

$$E_{\perp} = e_{n1} e^{ik_{n1}(z-z1)} + \sum_{n1=1}^{N} S_{11}(n1,1)\, \bar{e}_{n1} e^{ik_{n1}(z-z1)} \quad (7)$$

<u>Region 2 electric field</u>

$$S_{21}^{0} = T_z \left[ U - S_{22}^{I} T_g S_{11}^{II} T_g \right]^{-1} S_{21}^{I} \quad (8)$$

$$S_{11}^{0} = T_{-z} \left[ U - T_g S_{11}^{II} T_g S_{22}^{I} \right]^{-1} T_g S_{11}^{II} T_g S_{21}^{I}$$

$$S_{11}^{0} = T_{-z} \left[ U - T_g S_{11}^{II} T_g S_{22}^{I} \right]^{-1} T_g S_{11}^{II} T_g S_{21}^{I} \quad (9)$$

$$E_{\perp} = \sum_{n_2=1}^{N} \left( T_z S_{21}^{0} + T_{-z} S_{11}^{0} \right) \bar{e}_{n_2} \quad (10)$$

$$E_{\parallel} = \sum_{n_2=1}^{N} \left( T_z S_{21}^{0} - T_{-z} S_{11}^{0} \right) \bar{e}_{n2} Y_{n_2} \quad (11)$$

<u>Region 3 electric field</u>

$$E_{\parallel} = \sum_{n_3=1}^{N} S_{21}(n_3,1)\, e_{n_3} e^{-ik_{n3}(z-z2)} Y_{n_3} \quad (12)$$

$$E_{\perp} = \sum_{n_3=1}^{N} S_{21}(n_3,1)\, e_{n_3} e^{-ik_{n3}(z-z2)} \quad (13)$$

To verify the essential behaviour of the fields a wide-narrow-wide (WNW) junction is considered. A 2D field plot showing the radial dependence of the field is presented in Fig. 2. This simulation is very efficient as it required no more than a few seconds of CPU time in a PC compared to up to an hour on a comparable PC with a conventional numerical finite element code.

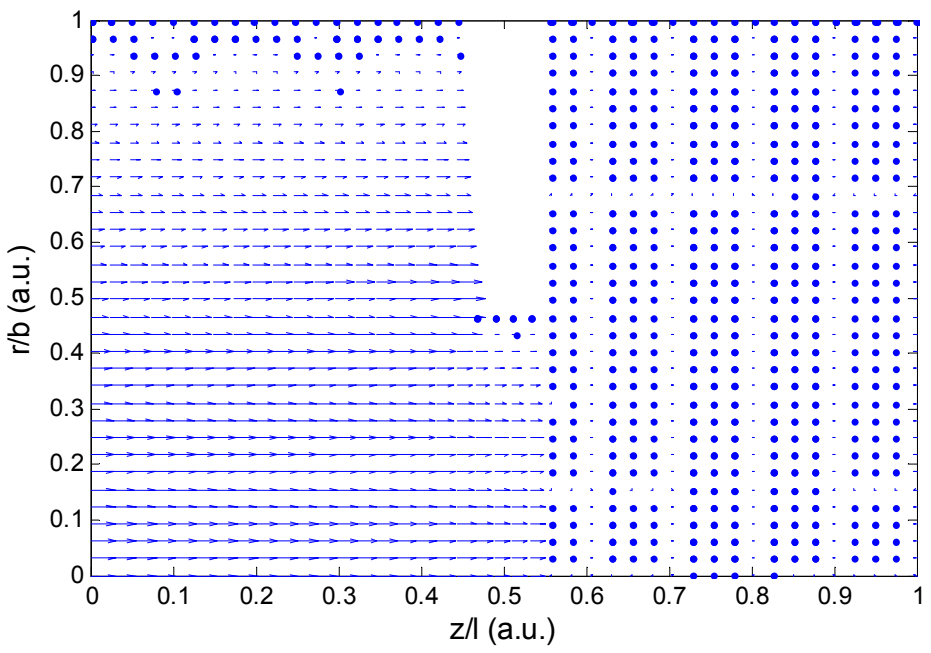

Figure 2: Axial electric $E_z$ for WNW transition as a function of radius and axial position.

## ANALYTICAL FORMULAE FOR DISPERSION CURVES

Combining GSM and mode matching in conjunction with the Bethe hole coupling perturbation theory [5,13] allows exact analytical expressions to be obtained for the normalised susceptance $\bar{B}$ of a structure. We note that $\bar{B}$ is equivalent to the coupling constant κ or fractional bandwidth of the mode. Let us consider the case of a monopole accelerating mode. From circuit theory [14] we can express the relationship between angular frequency ω and phase advance per cell Φ in the thin iris approximation (where κ<<1 ) as:

$$\omega = \omega_{\pi/2} \left( 1 - \frac{\kappa}{2} \cos\Phi \right) \quad (14)$$

where $\omega_{\pi/2}/2\pi$ is frequency as a phase advance of π/2. The relationship between phase and susceptance can be derived using either transmission line theory or from considering the ABCD matrix representation [4]:

$$\cos\Phi = \cos(k_z l) - \frac{\bar{B}}{2} \sin(k_z l) \quad (15)$$

In Eq. 15 l is the cavity length and $k_z$ is the propagation constant of the dominant mode: $k_z = \sqrt{(\omega/c)^2 - (\chi_{01}/b)^2}$ (where b is the radius in the cavity region and $\chi_{01}$ is the first root of the Bessel function $J_0$). Provided $k_z l << 1$ then to first order in $k_z l$ Eq. 15 is simplified to:

$$\cos\Phi = 1 - \frac{\bar{B}}{2} k_z l \quad (16)$$

A Bethe hole coupling analysis [5, 13] allows the fractional bandwidth to be obtained as:

$$\kappa = \frac{4a^3}{3\pi l b^2 J_1(\chi_{01})} \quad (17)$$

In Eq. 17 $a$ is the iris radius. Rewriting and comparing Eqns 15 and 16 in terms of $\cos\Phi$ allows the following equation to be derived in the thin iris approximation:

$$\bar{B} = \frac{3\pi b^4 J_1(\chi_{01})^2 k_z}{2a^3 \chi_{01}^2} \quad (18)$$

An exact formulation for the dominant mode can also be obtained by considering a susceptance placed symmetrically and in parallel with a transmission line. The susceptance normalised to the transmission line impedance is then obtained in terms of the reflection coefficient of $S_{11}(1,1)$ the dominant mode: [4]:

$$\bar{B} = \frac{2iS_{11}(1,1)}{S_{11}(1,1)+1} \quad (19)$$

The dispersion curve is obtained from the combination of Eqns. 15 and 18 have a limited domain of applicability due to the thin iris approximation used in obtaining it. An alternative method which circumvents this limitation is to consider the S matrix of a cell in an infinite periodic structure (obtained from GSM):

$$\begin{pmatrix} b_1 \\ b_2 \end{pmatrix} = \begin{pmatrix} S_{11} & S_{12} \\ S_{21} & S_{22} \end{pmatrix} \begin{pmatrix} a_1 \\ a_2 \end{pmatrix} \quad (20)$$

In Eq. 20 a represents the incident waves and b the reflected waves on each port, the subscripts refer to the ports. Applying the Floquet condition:

$$b_2 = e^{-i\Phi} a_1 \quad (21)$$

$$b_1 = e^{-i\Phi} a_2 \quad (22)$$

Eqns 20-22 represent an eigen-mode problem. Provided the field is evaluated sufficiently far away from the junction such that all evanescent modes have decayed to leave only the dominant propagating mode, we can solve for the phase advance. The solution for $\Phi$ in terms of the dominant propagating mode scattering matrix parameters is obtained:

$$\cos\Phi = \frac{1 + S_{21}(1,1)^2 - S_{11}(1,1)^2}{2S_{21}(1,1)} \quad (23)$$

The analytical equation for obtaining a dispersion curve using the thin iris approximation (Eqns 15 and 18), reveals there is little discrepancy between the two methods provided the iris is sufficiently thin (see Fig 3). However for irises thicker than t=50 μm the Bethe hole coupling formula that the analytical susceptance formula, which Eq 18 was based on, is no longer valid.

The analytical equation derived by considering the S matrix of a cell in an infinitely periodic structure (Eq. 23) is of course not limited by the thin iris approximation. However, it does require all evanescent modes to be small compared to the dominant mode. A Brillouin diagram is displayed in Fig 4 from a calculation based on three methods. The transverse mode matching method developed herein is in good agreement with results based on the longitudinal mode matching code KN7C [15,16]. However, there is some discrepancy between HFSS and the other methods. The bandwidth obtained from all three methods is: 1.074%, 1.100% and 1.066% for KN7C, analytical model and HFSS, respectively.

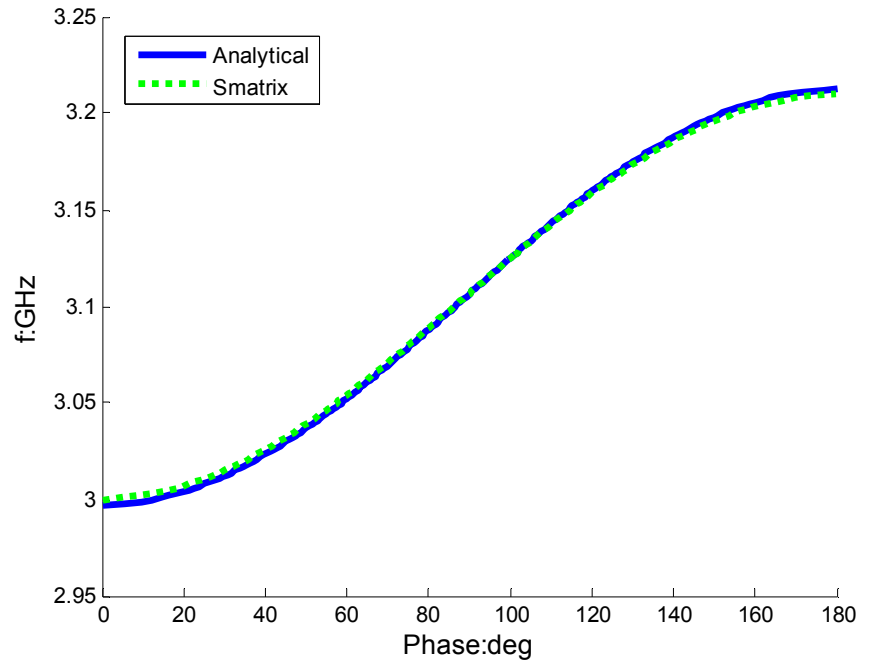

Figure 3: Brillouin diagram calculated using Eq. 15 for a=7.5mm, b=38.23mm, t=0.10mm, l=6.1mm. The analytical approximation corresponding to Eq. 18 is given in blue together with the exact model provided by Eq. 19 in green.

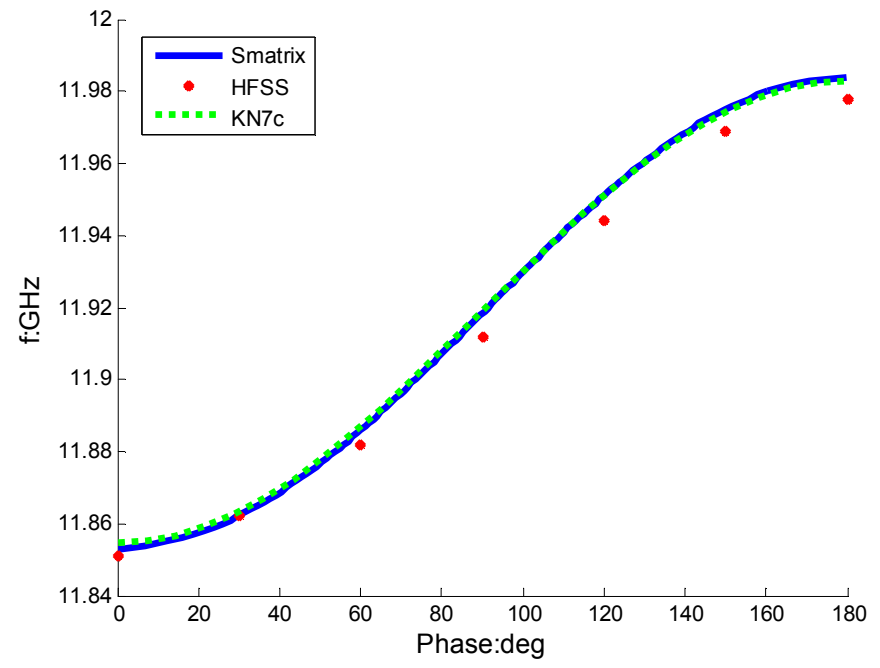

Figure 4 Brillouin diagram calculated using an analytical (Eq 23 blue) and numerical approaches (HFSS red and KN7c green) for an idealised CLIC_G cell. This structure has been represented as a simplified NWN transition with cell parameters: a=2.619mm, b=9.783mm, t=1.202mm, l=8.332mm.

## DISCUSSION

The transverse mode matching method developed herein provides an efficient and accurate means of characterising the fields and dispersion relation of accelerator structures. The analytical dispersion relation and analytic formula for iris suspectance can be applied to specific accelerator cells. However, the geometrical dependence of the analytical formula provides a rapid means of designing accelerator cells with sufficiently thin irises. It is planned to extend the mode matching method to consider aggregate accelerator structures.